\documentclass{raa}            % referee version: for submission

\usepackage{graphicx,times}             %for PS/EPS graphics inclusion, new
\usepackage{natbib}
\usepackage{amssymb,amsmath}
\usepackage{multirow}
\usepackage{booktabs}
\usepackage{url}

\bibpunct{(}{)}{;}{a}{}{,}
\newcommand{\tabincell}[2]{\begin{tabular}{@{}#1@{}}#2\end{tabular}}

\usepackage[a4paper=true,pagebackref=true]{hyperref}    %remove ',dvipdfm=false' by author
\hypersetup{colorlinks = true, linkcolor = green, anchorcolor = red, citecolor = blue, filecolor = red, pagecolor = red, urlcolor = red}

\begin{document}

   \title{Fine-grained Distributed Averaging for Large-scale Radio Interferometric Measurement Sets
%\footnotetext{$*$ Supported by the National Natural Science Foundation of China.}
}
%   \subtitle{I. Place Your Subtitle Here}

   \volnopage{Vol.0 (20xx) No.0, 000--000}      %%preserved for Editor. DOn't remove!
   \setcounter{page}{1}          %%starting page, preserved for Editor. DOn't remove!

   \author{ 
       Shoulin Wei\inst{1}
       \and Kaida Luo\inst{1}
       \and Feng Wang\inst{1,2}
       \and Hui Deng\inst{2}
       \and Ying Mei\inst{2}
   }

   \institute{Faculty of Information Engineering and Automation, Computer Technology Application Key Lab of Yunnan Province, Kunming University of Science and Technology, Kunming 650500, China;\\
        \and
            Center for Astrophysics, Guangzhou University, Guangzhou 510006, China; {\it fengwang@gzhu.edu.cn}\\          
\vs\no
   {\small Received~~2020 March 19; accepted~~2020~~September 22}}

\abstract{
%% --------------------------------------abstract
The Square Kilometre Array (SKA) would be the world’s largest radio telescope with eventually over a square kilometre of collecting area. However, there are enormous challenges in its data processing. The using of modern distributed computing techniques to solve the problem of massive data processing in SKA is one of the most important challenges. In this study, basing on the \textit{Dask} distribution computational framework, and taking the visibility function integral processing as an example,  we adopt a multi-level parallelism method to implement distributed averaging over time and channel.
\textit{Dask Array} was used to implement super large matrix or arrays with supported parallelism. To maximize the usage of memory, we further exploit the data parallelism provided by \textit{Dask} that intelligently distributes the computational load across a network of computer agents and has a built-in fault tolerance mechanism. The validity of the proposed pattern was also verified by using the Common Astronomy Software Application (CASA), wherein we analyze the smearing effects on images reconstructed from different resolution visibilities.
%% --------------------------------------keyword
}

   \authorrunning{Wei et al. }            %author_head in even pages
   \titlerunning{Fine-grained Distributed Averaging for MS}  % title_head in odd pages

   \maketitle
%% The author head (on even pages) and the title head (on odd pages) will be
%% automatically extracted from \author{} and \title{}. Whenever the title is too long,
%% you will be asked to supply a shorter one by inserting either \authorrunning{} or
%% \titlerunning{} before \maketitle. Anyway, you can specify your own heads.
%%
%%
%% Note: In the following text body of your manuscript, please note several differences from
%%       other major journals:
%% (1) \subsection{Please Capitalize the First Letter of Each Notional Word in Subsection Title}
%% (2) Please Capitalize the First Letter of Each Notional Word in all tables' captions

%
%________________________________________________ sections below
%
\section{Introduction}           %% first-level sections will be auto-capitalized
\label{sect:intro}
The primary reason for constructing the Square Kilometre Array (SKA) (\citealt{dewdney2009square}) that comprises many individual stations is to explore the universe with the highest resolution, sensitivity and dynamic range. However, the telescope brings new scientific and technical challenges to the modern Information Technology. One of the biggest challenges is managing the extremely large data volumes generated by the SKA every day. The narrow channel bandwidth and short correlator timestep do minimize the decorrelation effects on the longest baselines but increase the amount of visibility data. This substantial increase in the bulk of data requires the development of new and sophisticated distributed data processing schemes and concurrent algorithms that can run on high-performance computing hardware.

To prompt the development of Science Data And Handling Processor (SDHP), the Radio Astronomy Simulation, Calibration and Imaging Library (RASCIL) (\citealt{RASCIL}), including radio interferometry calibration and imaging algorithms in Python and NumPy, has been developing for a few years. The interfaces all operate with familiar data structures such as image, visibility table, gain table, and so on. To achieve sufficient performance, the latest version of the RASCIL takes a dual pronged approach - using threaded libraries for shared memory processing, and the \textit{Dask} (\cite{rocklin2015dask}) library for distributed processing. 

\textit{Dask} is a flexible library for parallel computing in Python that composed of two parts, i.e., Dynamic task scheduling and Big Data Collection. In the area of interferometric data processing, \textit{dask-ms} (\citealt{daskms}) and \textit{cngi-prototype} (\citealt{cngi}) exploited \textit{Dask} arrays  to expose columns of tables in Measurement Sets (MS) (\cite{van2015casacore}) to support parallel operations on MS. Current RASCIL software takes full advantage of \textit{Dask}'s dynamic scheduling capabilities. However, the big data collection feature has never been tested in RASCIL development. Obviously, distributed computing is the unique choice for processing large-scale astronomical data. The visibility has time and bandwidth separation, making distributed computing a natural solution.

Averaging in post-correlation is considered to reduce the volume of astronomical data and computational cost (\cite{offringa2016compression}). To further investigate the ability of \textit{Dask} on its big data processing, we take the visibility function data averaging(integral) as the case and developed a fine-grained distributed averaging algorithms including time and channel averaging so as to test whether \textit{Dask} big data collection feature can be used in the subsequent version of the RASCIL.

The rest of this paper is organized as follows: Section~\ref{sect:avg}~introduces the theoretical analysis of averaging; Section~\ref{sect:implement}~describes the algorithm implementation and validation results; Section~\ref{sect:eval}~analyzes the performance evaluation in terms of granularity; and Section~\ref{sect:conclusion}~presents the conclusion. The code that replicates the results in this work is available online\footnote{\url{https://github.com/astroitlab/vis_averaging}}.

%% Authors can give a citation as 'Michel et al. 1992'.
%% You may also use \cite, \citep and \citet for citation, and use Table~1 or Figure~1
%% and so forth. Using \ref and \label for cross-references of Tables/Figures
%% is a good way in adjusting/adding/removing text, tables or figures.

\section{Radio Interferometric Averaging}
\label{sect:avg}
This section provides a brief overview of radio interferometry, reviews the theory of averaging, and discusses the effects of averaging on interferometric data processing.

 The response of interferometric arrays to the brightness distribution of sources in the sky is measurements called \emph{visibilities},  which are complex values sampled from the Fourier transform of the sky brightness distribution (\cite{Thompson1999}). The relation between the synthesis array, visibilities, and sky brightness is well established, and it can be expressed by the radio interferometer measurement equation (RIME) (\citealt{hamaker1996understanding}). The RIME has provided a rigorous mathematical framework to model the signal propagation effects and calculate the visibilities. We assume that a baseline is composed of two spatially separated stations $p$ and $q$. The visibility matrix $\boldsymbol{V}_{pq}$ is calculated as an integral product over a time and frequency intervals [$t_0$, $t_1$] $\times$ [$\nu_0$, $\nu_1$]. It can be expressed as
\begin{equation}
    <\boldsymbol{V}_{pq}> = \lim_{\Delta t,\Delta \nu \to 0} \frac{1}{\Delta t \Delta \nu} \int_{t_0}^{t_1} \int_{\nu_0}^{\nu_1} \boldsymbol{J}_p(t,\nu)~\boldsymbol{B}~\boldsymbol{J}_q^H(t,\nu)~d\nu~dt
\label{eq:rime}
\end{equation}
The angled brackets denote data averaging over one baseline, frequency channel, and integration timestep. In addition, $H$ represents the conjugate transpose. Matrix $\boldsymbol{B}$ describes the continuous brightness distribution, and $\boldsymbol{J}$ describes the Jones chain modeling effects along the signal propagation path (\cite{Smirnov2011a}).

\subsection{Channel Averaging}
 Modern radio telescopes typically make observations across multiple channels, each of which is bound to a specific frequency. Channel averaging is used to integrate the visibilities across several continuous or discrete channels. To keep both sides of the equation’s form consistent, the weighted average visibility across channels can be expressed as
\begin{equation}
\boldsymbol{V}_{pq}^{avg}~\boldsymbol{W}_{pq}^{avg}=\sum_i^{nch}~\boldsymbol{V}_{pqi}~\boldsymbol{W}_{pqi}
\label{eq:eqform}
\end{equation}
The weight of the broadened channel is the sum of the weights of the integral channels. All channels have the same weight at a specific time and baseline. Thus,
\begin{equation}
\boldsymbol{W}_{pq}^{avg}=\sum_i^{nch}~\boldsymbol{W}_{pqi}=nch \times \boldsymbol{W}_{pqi}
\label{eq:weight}
\end{equation}
where $i$ is the index of the frequency channel, $nch$ is the number of integral channels, and $avg$ denotes the channel averaging operation. In addition, $\boldsymbol{V}$, $\boldsymbol{W}$ represent the columns of DATA and WEIGHT columns in the MS, respectively (as shown in Table~\ref{Tab:mslayout}). For brevity, the angle brackets of $\boldsymbol{V}$ in Eq.~(\ref{eq:eqform}) are omitted. Substituting Eq.~(\ref{eq:weight}) into Eq.~(\ref{eq:eqform}) gives the integral visibilities over the $nch$  channels finally as the weighted average

\begin{equation}
\boldsymbol{V}_{pq}^{avg}=\frac{\sum_i^{nch}~\boldsymbol{V}_{pqi}~\boldsymbol{W}_{pqi}}{\Delta \theta+nch \times \boldsymbol{W}_{pq}^{avg}}
\label{eq:avgchannel}
\end{equation}

To avoid division by zero, we introduce a value closer to zero $\Delta \theta$ in Eq.~(\ref{eq:avgchannel}). The effects of $\Delta \theta$ are discussed in Section~\ref{subsec:accur}.

%
%               one-column-spanning table
\begin{table}[h]
\begin{center}
\caption[]{ MS Sectional Layout Referenced in this Work.}\label{Tab:mslayout}
%%Please Capitalize the First Letter of Each Notional Word in table's caption
 \begin{tabular}{clclc}
  \hline\noalign{\smallskip}
Columns & Types $|$ Dimensional transformation & Comments              \\
  \hline\noalign{\smallskip}
UVW & double $|$ ($N_r$, 3)$\to$ ($N_t$, $N_b$, 3) & $uvw$ coordinates units in meters \\
DATA\_DESC\_ID & int $|$ ($N_r$,) &	Data description identifier ($\ge$ 0) \\
FIELD\_ID & int $|$ ($N_r$,) &	Field identifier ($\ge$ 0)    \\
WEIGHT & float $|$ ($N_r$, $N_c$)$\to$($N_t$, $N_b$, $N_c$) & Weight for whole data matrix   \\
DATA & complex $|$ ($N_r$, $N_f$, $N_c$)$\to$($N_t$, $N_b$, $N_f$, $N_c$) & Complex visibilities matrix   \\
\noalign{\smallskip}\hline
\end{tabular}
\end{center}
Notes~~~~$N_r$ = number of rows in MS table, $N_f$ = number of frequency channels, $N_c$ = number of correlators $N_t$=number of timesteps, and $N_b$=number of baselines. Dimension refers to the shape of the entire specified column rather than that of a single cell.
\end{table}

\subsection{Time Averaging}

Assume that $t$ time points are sampled in an averaging period. In terms of the averaging time, these t sampled points will be integrated into one time point that includes the visibilities of n(n-1)/2 baselines. In other words, the number of visibilities is reduced from $t \times n \times (n-1)/2$ to n(n-1)/2. The averaged weights and visibilities are given by
\begin{equation}
\boldsymbol{W}_{pq}^{avg}=\sum_j^{t}~\boldsymbol{W}_{pqj}
\label{eq:t_weight}
\end{equation}

\begin{equation}
\boldsymbol{V}_{pq}^{avg}=\frac{\sum_j^{t}~\boldsymbol{V}_{pqj}~\boldsymbol{W}_{pqj}}{\Delta \theta+\boldsymbol{W}_{pq}^{avg}}
\label{eq:t_vis}
\end{equation}
Here, $j$ is the index of sampled time point. As shown in Eq.~(\ref{eq:t_weight}) and Eq.~(\ref{eq:t_vis}),  the averaged weights and visibilities for time averaging have a similar structure to those for channel averaging. The difference is that time averaging integrates data with the same baseline along the time axis. As time averaging combines different time points into one, the time-dependent data also needs to be recalculated. The time-dependent data including $U,V,W$ and the new weighted $Time$ are given by
\begin{equation}
\Big{(}
\boldsymbol{U}=\frac{\sum_j^{t} U_j \overline{\boldsymbol{W}_{j}}}{\sum_j^{t} \overline{\boldsymbol{W}_{j}}},
\boldsymbol{V}=\frac{\sum_j^{t} V_j \overline{\boldsymbol{W}_{j}}}{\sum_j^{t} \overline{\boldsymbol{W}_{j}}},
\boldsymbol{W}=\frac{\sum_j^{t} W_j \overline{\boldsymbol{W}_{j}}}{\sum_j^{t} \overline{\boldsymbol{W}_{j}}} 
\Big{)}
\label{eq:t_uvw}
\end{equation}

\begin{equation}
\boldsymbol{Time}=\frac{\sum_j^{t} T_j \overline{\boldsymbol{W}_{j}}}{\sum_j^{t} \overline{\boldsymbol{W}_{j}}}
\label{eq:t_time}
\end{equation}

Here, $\overline{\boldsymbol{W}_{j}}$ represents the mean weight across all of the baselines at one time point.

\subsection{Averaging Effects}
The measurements at each baseline inevitably contain uncorrelated Gaussian noise that creates uncertainty in the real and imaginary parts of the visibilities. This uncertainty is mainly represented by signal fluctuation. Eq.~(\ref{eq:avgchannel}) and Eq.~(\ref{eq:t_vis}) are essentially weighted averages, where the smoothing effect can suppress the fluctuation of visibilities. Moreover, multiple consecutive channels or timesteps are integrated into a broader channel or period that can significantly reduce the volume of visibilities. However, this form of data compression unavoidably results in a net loss of amplitude, known as smearing (or time/bandwidth decorrelation in the $uv$-plane), which mainly manifests a decrease in the amplitude toward off-center sources. The amplitude reduction in source flux increases with baseline length and distance from phase center (\citealt{wijnholds2018baseline}). Smearing also has detrimental effects on sensitivity, as the noise amplitude does \emph{not} decrease. If we consider the entire $uv$-plane, smearing distorts the point spread function (PSF) and produces relatively high sidelobes (second column in Figure~\ref{fig:ms2020-0112fig3}), as the weighted average results in coarser time/frequency bins or lower resolution re-samples. Furthermore, the wide field effects of smearing are of particular concern in the context of modern radio interferometers, and in this case, some bright sources far from the FOV that are modulated by PSF sidelobes can contribute global background noise and artifacts to the image (\citealt{smirnov2012understanding}).

\section{Implementation}
\label{sect:implement}
This section describes the architecture of the proposed method and illustrates the principles of different granularity distributions. The accuracy of the proposed method is explained in comparison to those using the common astronomy software application (CASA) (\citealt{mcmullin2007astronomical}) software.

\subsection{Architecture}
\begin{figure}[h]
    \centering
    \includegraphics[width=14.0cm]{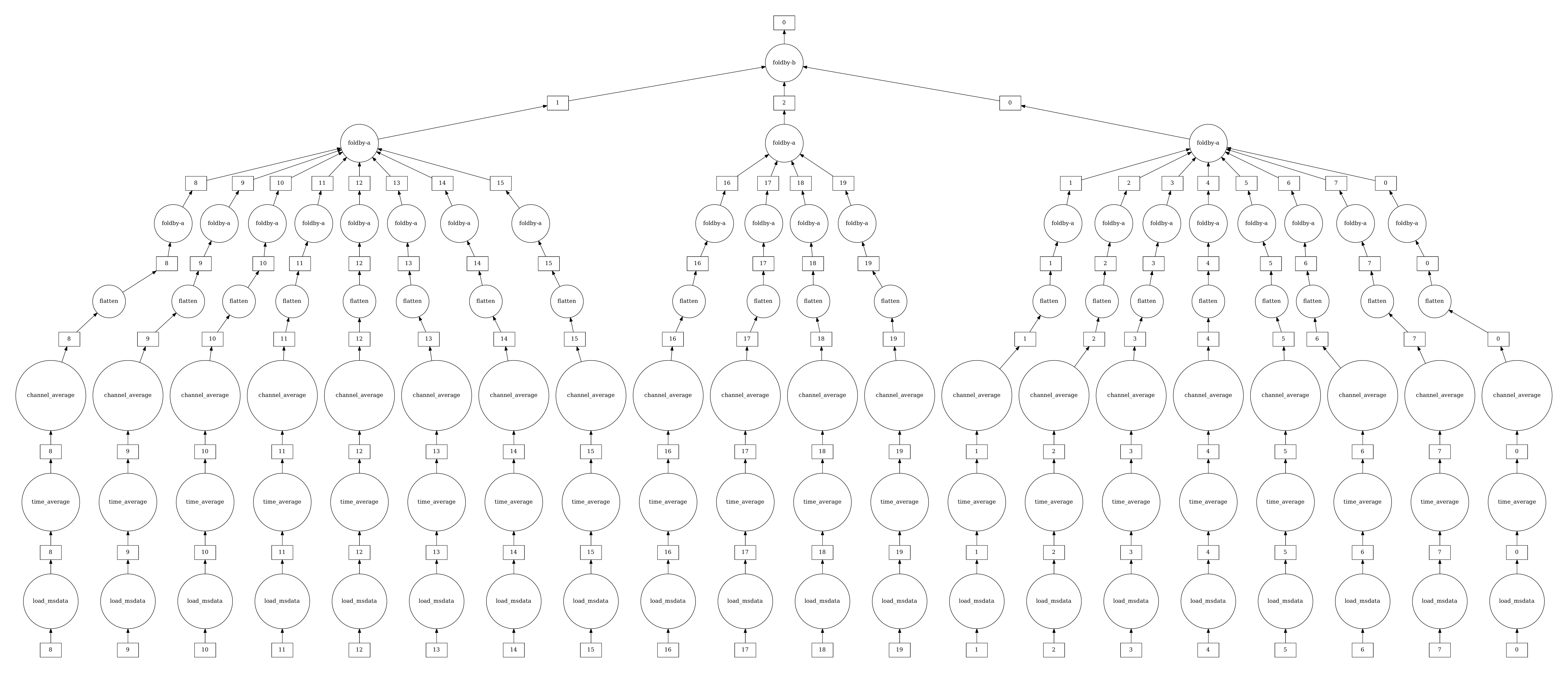}
    \caption{Dask task graph for averaging computation across 20 measurement sets.}
    \label{fig:ms2020-0112fig1}
\end{figure}

The key goal of this study is to allow averaging to run in a parallel and distributed manner. \textit{Dask} was selected for this task owing to its flexibility and limited invasion into the existing codes. In addition, \textit{Dask} provides much of the NumPy application programming interface that users are familiar with, which means simple NumPy-style iterative optimization algorithms can be written that will leverage the built-in parallel data model. An averaging computation across 20 MSs is visualized, as shown in Figure \ref{fig:ms2020-0112fig1}, and generated using \textit{Dask}’s built-in visualization features.

\begin{figure}[h]
    \centering
    \includegraphics[width=0.95\textwidth]{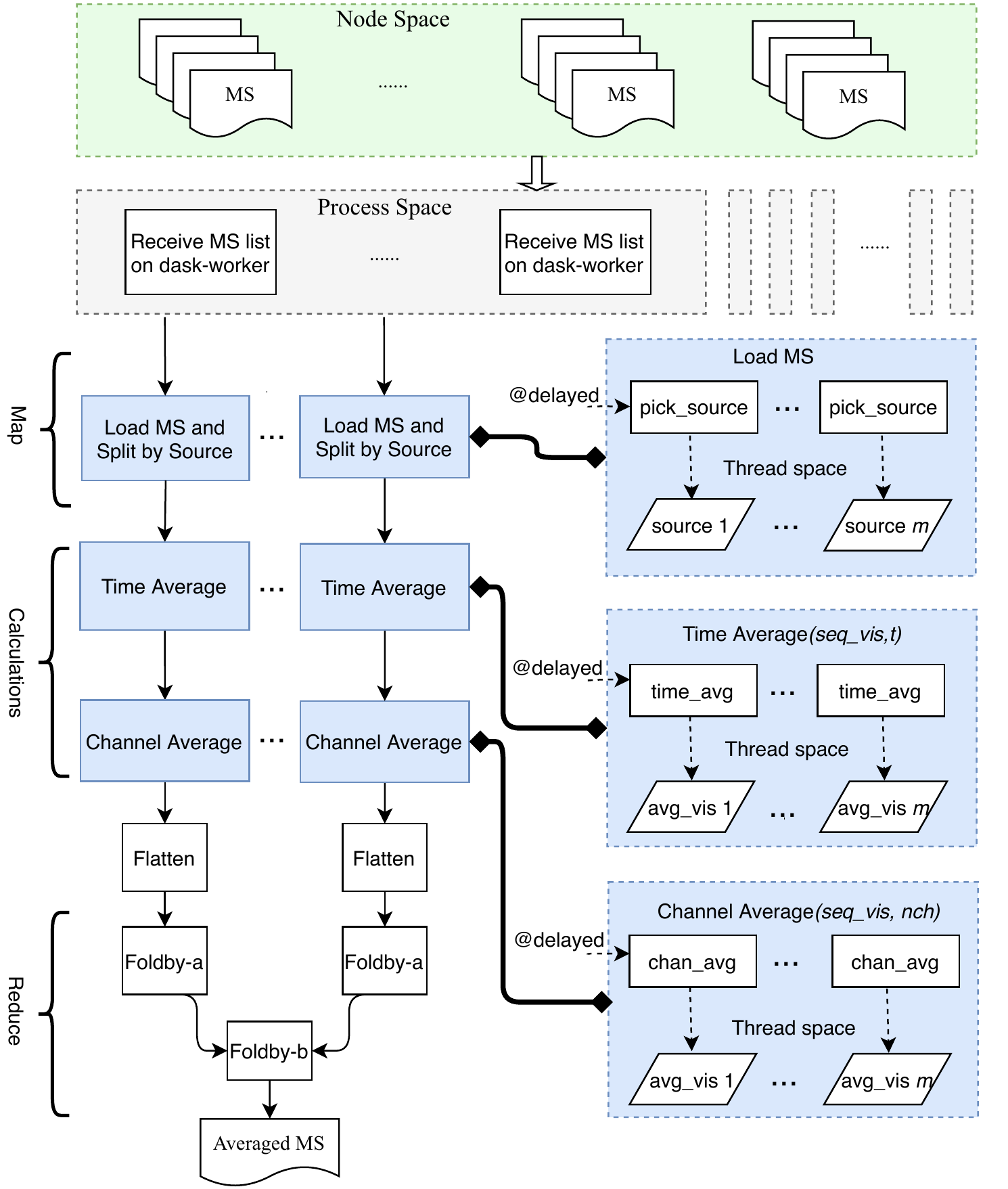}
    \caption{Multi-level parallelism architecture starts with the MS list loading in the node space. When a node receives a partial list, it begins to load the MS and splits it by source and subsequent operations. This is shown by the solid blue boxes that link to the corresponding dashed blue boxes to elaborate the internal execution.}
    \label{fig:ms2020-0112fig2}
\end{figure}

To provide clarity with respect to the various levels of parallelism, Figure \ref{fig:ms2020-0112fig1} was abstracted into Figure \ref{fig:ms2020-0112fig2}. In accordance with the requirements of computing resources, various logical parts of the program are arranged across different executing spaces, as shown in Figure \ref{fig:ms2020-0112fig2}. 

Visibilities and related configurations are commonly stored as an MS in a folder. The process starts with the MS loading from a shared file system mounted to the same path in each node. Multiple MS files are loaded in parallel cross nodes. In a single node, an MS is loaded and split by FIELD\_ID and DATA\_DESC\_ID in parallel within one process, where each dask worker gets its own interpreter and memory space for computation. Python global interpreter lock (GIL) only allows one thread to control the Python interpreter at a time, which can be a bottleneck that prevents full multi-core usage. Therefore, multi-process can bypass the GIL to allow multiple cores to execute in parallel. 

Throughout its architecture, this algorithm adopts the map/reduce strategy for data analysis. In the map phase, each MS is secondary split by source. In particular, UVW (Table~\ref{Tab:mslayout}) is not converted from meters to wavelengths, otherwise the UVW volume would increase. Multi-level parallelism and parallel multi-dimensional arrays (discussed in Section \ref{sect:datamodel}) can achieve fine granularity of parallel averaging on visibilities. In the reduction process, all averaged visibilities are merged based on the source attribute. However, combination inevitably introduces a time penalty due to the huge amount of data between workers. This study adopted the high-order function \textit{Bag.foldby}, providing efficient parallel aggregation with minimal communication.

\subsection{Data Structure Definition}
\label{sect:datamodel}
The MS format is widely used by modern telescopes because it designs to avoid data redundancy and provides optional sub-tables/columns for universality. We utilize the \textit{python-casacore} (\cite{van2015casacore}), a python interface binding to the casacore library compiled by C++ in radio astronomy, to read and modify the MS. In this study, \textit{dask.bag} to wrap the list of MS paths. When mapping on \textit{dask.bag}, the dask scheduler dispatch splits of \textit{dask.bag} between various workers. Averaging includes complex array computations across the DATA, WEIGHT, and UVW columns that contain most of the MS data. Here, \textit{dask.array} was used to load these three columns as it allows the arrays to be split into chunks and processed in parallel. In addition, the averaging calculation refers to other configurations such as the antennas, polarization, and baseline positions. XArray was used to manage multiple arrays and relate the configurations as a consistent dataset. Table \ref{Tab:xarray_model} lists the XArray dataset definitions for the experimental data \footnote{The data can be downloaded from \url{ http://casa.nrao.edu/Data/EVLA/IRC10216/day2_TDEM0003_10s_norx.tar.gz} (the data size is 1.1 GB, the extracted data is about 2.0 GB).}.
%               one-column-spanning table
\begin{table}[h]
\begin{center}
\caption[]{XArray Dataset Definitions for the Experimental Data.}\label{Tab:xarray_model}
%%Please Capitalize the First Letter of Each Notional Word in table's caption
 \begin{tabular}{ll}
  \hline\noalign{\smallskip}
\textbf{Dimensions:}      &        (nant: 19, nbaseline: 171, nchan: 64, nfpos: 3, npol: 4,  nspos: 3, ntime: 17)\\
\noalign{\smallskip}\hline
\textbf{Coordinates:}     &  \\
\hspace{2ex}  * ntime        &      (ntime) int64 0 1 2 3 4 5 6 7 8 9 10 11 12 13 14 15 16\\
\hspace{2ex}  * nbaseline    &      (nbaseline) $<$U5 '0-1' '0-2' '0-3' ... '16-18' '17-18'\\
\hspace{2ex}  * nchan        &      (nchan) int64 0 1 2 3 4 5 6 7 ... 56 57 58 59 60 61 62 63\\
\hspace{2ex}  * npol         &      (npol) $<$U2 'p0' 'p1' 'p2' 'p3'\\
\hspace{2ex}  * nfpos        &      (nfpos) $<$U1 'u' 'v' 'w'\\
\hspace{2ex}  * nant         &      (nant) int64 0 1 2 3 4 5 6 7 ... 11 12 13 14 15 16 17 18\\
\hspace{2ex}  * nspos        &      (nspos) $<$U1 'x' 'y' 'z'\\
\noalign{\smallskip}\hline  
\textbf{Data variables: }&  \\
\hspace{2ex}    vis          &      (ntime, nbaseline, nchan, npol) complex64 \\ & dask.array$<$chunksize=(17, 171, 64, 4), meta=np.ndarray$>$\\
\hspace{2ex}    weight       &      (ntime, nbaseline, npol) float64  dask.array$<$chunksize=(17, 171, 4), meta=np.ndarray$>$\\
\hspace{2ex}    uvw          &      (ntime, nbaseline, nfpos) float64 dask.array$<$chunksize=(17, 171, 3), meta=np.ndarray$>$\\
\hspace{2ex}    frequency    &      (nchan) float64 3.639e+10 3.639e+10 ... 3.64e+10 \\
\hspace{2ex}    channel\_bandwidth & (nchan) float64 1.25e+05 1.25e+05 ... 1.25e+05 1.25e+05 \\
\hspace{2ex}    xyz          &      (nant, nspos) float64 -1.602e+06 -5.042e+06 ... 3.555e+06 \\
\hspace{2ex}    mount        &      (nant) $<$U6 'ALT-AZ' 'ALT-AZ' ... 'ALT-AZ' 'ALT-AZ'\\
\hspace{2ex}    names        &      (nant) $<$U4 'ea01' 'ea02' 'ea03' ... 'ea25' 'ea27' 'ea28'\\
\hspace{2ex}    diameter     &      (nant) float64 25.0 25.0 25.0 25.0 ... 25.0 25.0 25.0\\
\noalign{\smallskip}\hline    
\textbf{Attributes: }    &  \\
\hspace{2ex}    tag:        &   0\_7\\
\hspace{2ex}    phasecentre: &  $<$SkyCoord (ICRS): (ra, dec) in deg    (202.78453327, 30....\\
\hspace{2ex}    polarisation: & [[5 6 7 8],[5 6 7 8]]\\
\noalign{\smallskip}\hline
\end{tabular}
\end{center}
\end{table}

\subsection{Accuracy}
\label{subsec:accur}
Due to instrumental failure, partial baselines may be missing ($N_b<$n(n-1)/2) from raw measurement data. Therefore, matrices may be a different shape when they are loaded into the memory. For convenience, the matrices are filled with zeros to compensate for inconsistencies in the data dimensions. Additionally, in Eq.~(\ref{eq:avgchannel}) and Eq.~(\ref{eq:t_vis}), a minimal value ($\Delta \theta$) was introduced into the denominators to avoid the problem of dividing by zero. However, $\Delta \theta$ also introduces accuracy loss. Lower value of $\Delta \theta$ equates to better accuracy. The fact that if two matrices ($\boldsymbol{A}$ and $\boldsymbol{B}$) are numerically close, the mean and variance of \emph{loss error} ($\boldsymbol{A}$-$\boldsymbol{B}$) between them will approach zero was used to examine the accuracy of our proposed method compared to the CASA software. As shown in Table \ref{Tab:accuracy}, the mean and variance of accuracy loss are close to zero, proving that the two visibility matrices are almost identical.
\begin{table}[h]
\begin{center}
\caption[]{Accuracy Loss of DATA Column with Time Intervals $t$ of 2 and Various Channels $nch$.}\label{Tab:accuracy}
%%Please Capitalize the First Letter of Each Notional Word in table's caption
\begin{tabular}{clclclclcl}
  \hline\noalign{\smallskip}
\multirow{2}*{\tabincell{c}{Averaged \\ parameters}} & \multicolumn{2}{c}{Mean of} & \multicolumn{2}{c}{Variance of} \\
\cmidrule(r){2-3} \cmidrule(r){4-5}
  & real part & imaginary part & real part & imaginary part \\
  \hline\noalign{\smallskip}
t=2,nch=2 & -2.21427233e-13 & -6.91339540e-13 & 4.25529084e-20 & 4.12462738e-20 \\
t=2,nch=4 &	-3.44819459e-13 & -1.11037129e-12 & 4.29342260e-20 & 4.11034429e-20  \\
t=2,nch=8 &	-9.18691002e-13 & 3.06048242e-14 &  5.25735201e-20 & 5.13628739e-20  \\
~~t=2,nch=16 & -6.10984707e-13 & -2.06883846e-13 &  8.46708015e-20 & 7.86904922e-20  \\
~~t=2,nch=32 &  6.62003601e-13 & 1.14721947e-12 & 1.42577666e-19 & 1.30641657e-19  \\
~~t=2,nch=64 & -5.08093825e-12 & -7.35946702e-12 & 2.92768210e-19 & 2.52621540e-19 \\
\noalign{\smallskip}\hline
\end{tabular}
\end{center}
Notes~~~~The experimental MS was averaged from the original 1-s correlator visibility integration time to 10-s averages; thus, $t$=2 represented 20-s time averaging. The averaged visibilities were calculated using $\Delta \theta$ = 1e-16 in our programs and the \textit{split} task in CASA with parameters $timebin$='20s' and $width$=$nch$, respectively. 
\end{table}

\subsection{Compression Ratios and Imaging Artifacts}
\begin{figure}[h]
  \begin{minipage}[t]{0.499\linewidth}
  \centering
   \includegraphics[width=0.95\linewidth,trim=50 370 100 30,clip]{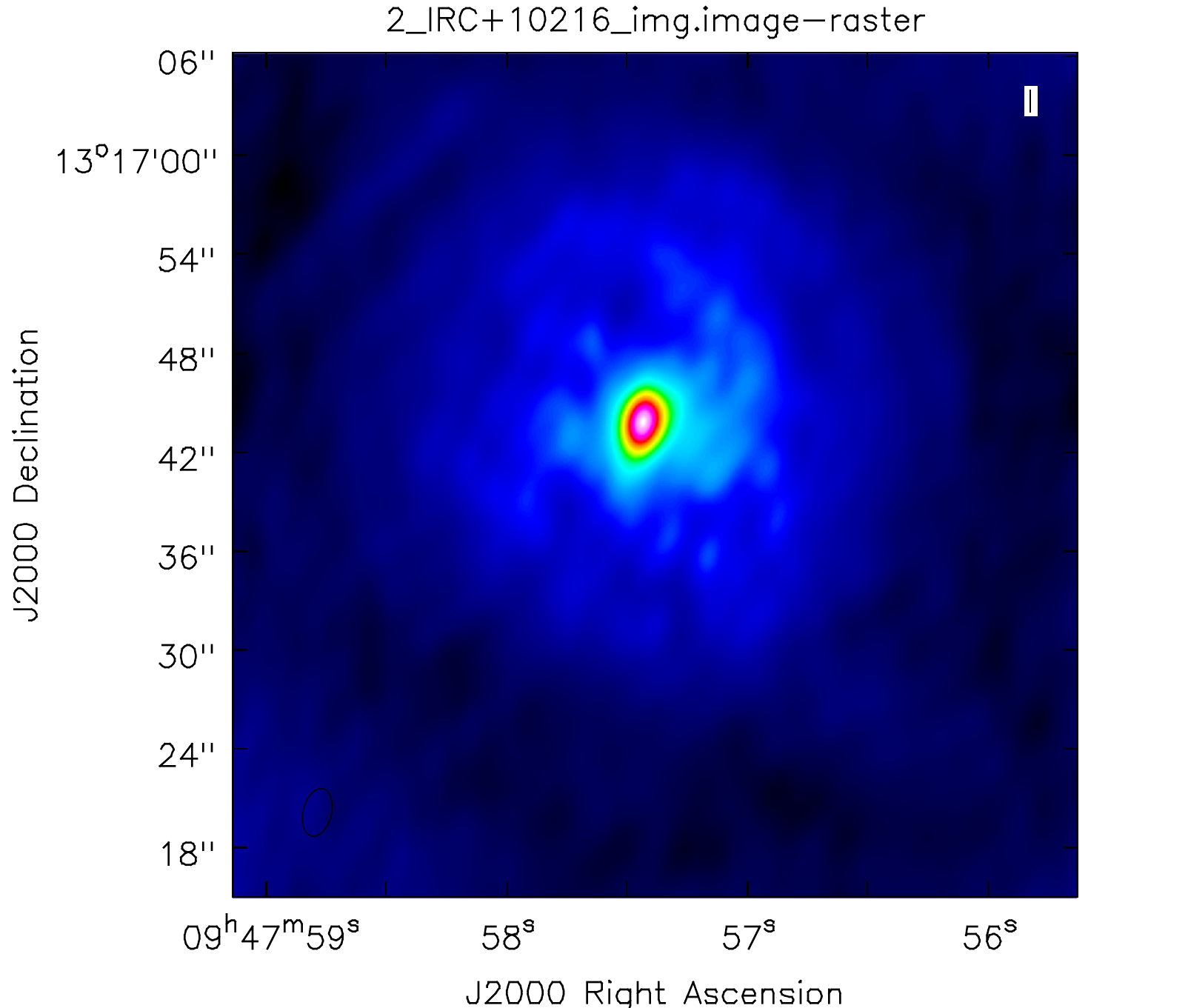}%trim left down right up
   \subcaption{}
  \end{minipage}%
    \begin{minipage}[t]{0.499\linewidth}
  \centering
   \includegraphics[width=0.95\linewidth,trim=50 370 100 30,clip]{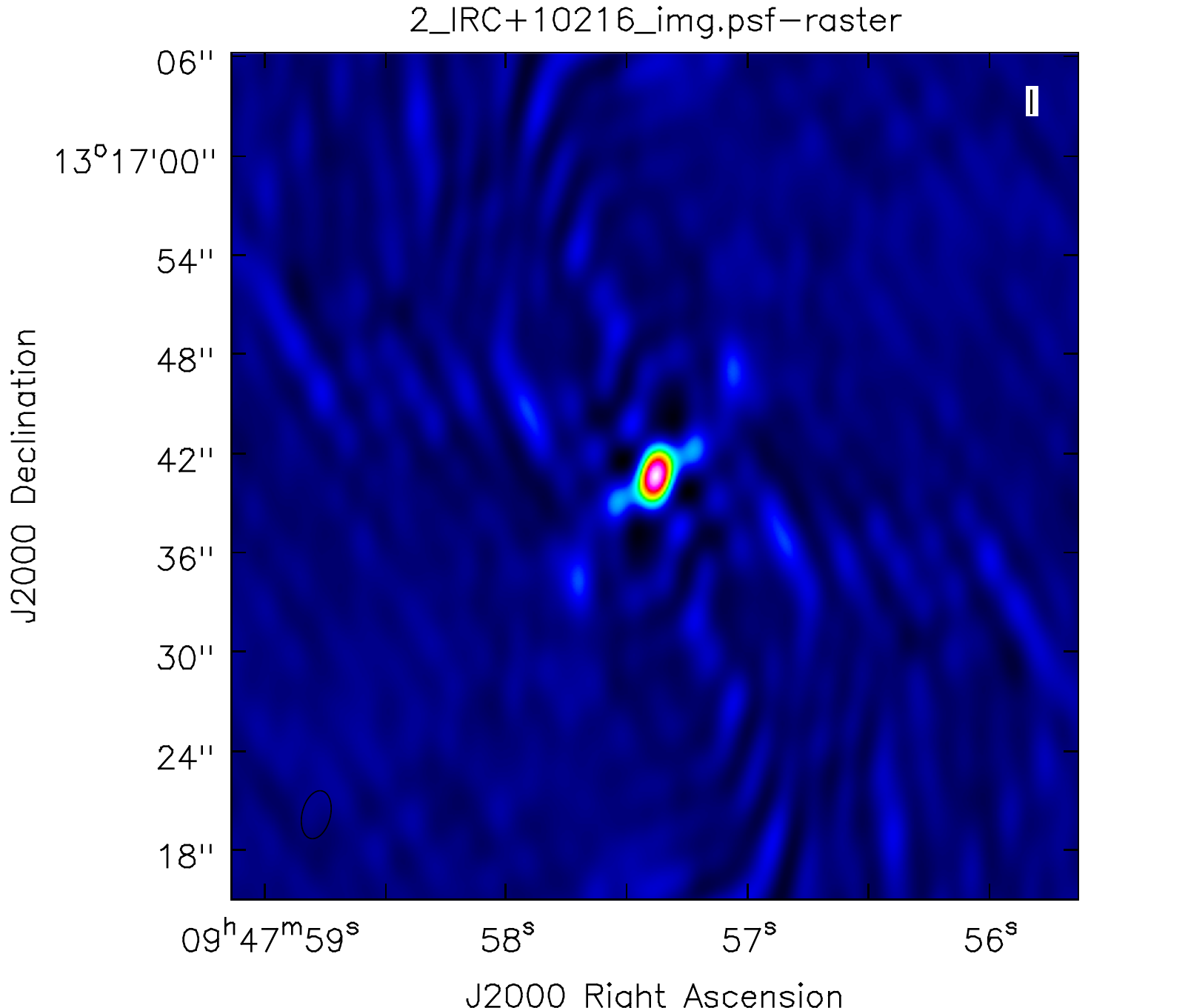}%trim left down right up
   \subcaption{}
  \end{minipage}%  
  
\begin{minipage}[t]{0.499\textwidth}
  \centering
   \includegraphics[width=0.95\linewidth,trim=50 370 100 30,clip]{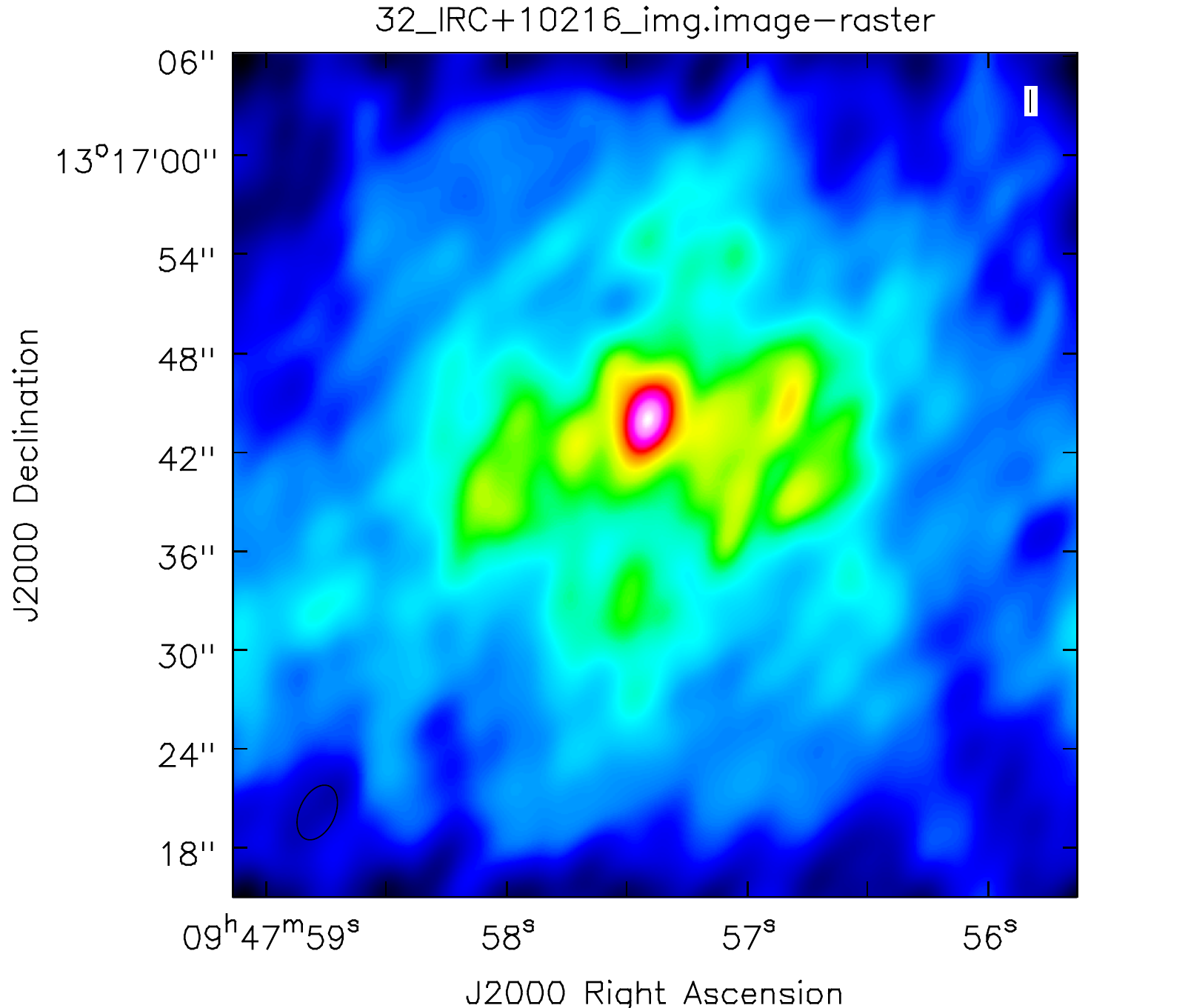}
   \subcaption{}
  \end{minipage}%
\begin{minipage}[t]{0.499\linewidth}
  \centering
   \includegraphics[width=0.95\linewidth,trim=50 370 100 30,clip]{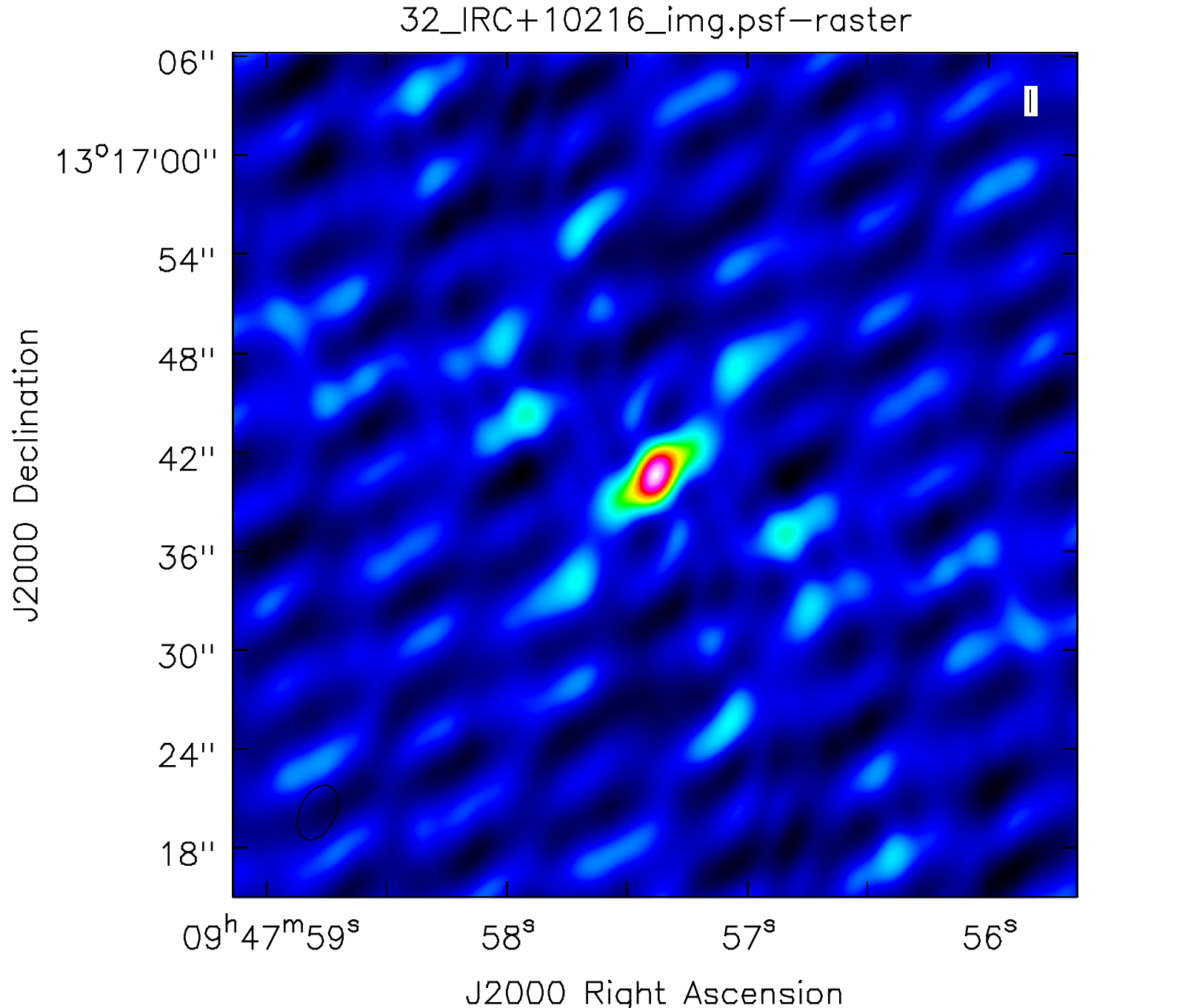}%trim left down right up
   \subcaption{}
  \end{minipage}%
  \caption{Images corresponding to PSFs produced under the same processing flows with different integral channels, $nch$=2 (Figure (a),(b)) and $nch$=32 (Figure (c),(d)). Excessive averaging over channels may cause the image to become blurry.}
  \label{fig:ms2020-0112fig3}
\end{figure}

Theoretically, data can be averaged over several timesteps and channels with the arguments $t$ (default:2) and $nch$ (default:4). The compression ratio (CR) is the size of the unaveraged visibilities to the averaged one (and thus CR = $t \times nch$ in this study). \cite{cotton2009} has qualitatively analyzed the CR for time averaging and the level of artifacts in the context of the VLA. He observes that time averaging cannot be of an arbitrary length; otherwise, excessive amplitude attenuation would reduce the \textit{dynamic range} (= ratio of the peak flux to off-center source) in the vicinity of bright sources and produce artifacts that degrade the quality of the image (for detailed image artifacts caused by time smearing, see Figure 4 in \cite{cotton2009}). Moreover, if the phase variation as a function of channel is large, it may be necessary to restrict the averaging to a few channels to prevent the introduction of delay-based closure errors, which can be caused by averaging over non-bandpass corrected channels with large phase variations. To explain frequency smearing, the channel-averaged data were rewritten back to the original MS, and the source images and PSFs were produced (Figure ~\ref{fig:ms2020-0112fig3}) by the \emph{tclean} task in CASA. The experimental data were obtained by observing two spectral lines produced by the asymptotic giant branch star IRC+10216, where the dense inner envelope (for more information about the circumstellar envelope, see \citealt{leao2006circumstellar}) produced the SiS line as a centralized emission, and the external envelope produced the HC3N line as a ring of emission. The HC3N can re-radiate energy absorbed from the SiS radiation.

Time and channel averaging on visibilities can be treated as a form of \textit{pseudo-convolution} by some ensemble boxcar-like window functions (\citealt{atemkeng2016using}), and the sizes of boxcar are proportional to the $uv$-extent of the time and frequency bins (related to the arguments $t$ and $nch$). The Fourier transform of a boxcar-like is a sinc-type taper, and the scale of the tapering response is inversely proportional to the scale of the window function. Larger arguments result in wider boxcars corresponding to narrower tapers with higher sidelobes that are therefore more likely to smearing. As shown in Figure~\ref{fig:ms2020-0112fig3}, frequency smearing causes a radial broadening distortion of the PSF. Thus, the amplitude of the SiS, modulated by the high sidelobes, translates into an unwanted background in the domain of the HC3N.

\section{Performance Evaluation}
\label{sect:eval}

This section explores the performance characteristics of the algorithm. All of the experiments were performed on ten Ubuntu 18.04 server workstations, two of which are equipped with 56 processors (Intel Xeon CPU E5-2660 v4), hyperthreading, 3.4-GHz maximum frequency, and 512 GB of RAM (i.e., each 32 GB on 16 sockets) and the rest of which equipped with 32 processors (Intel Xeon CPU E5-2630 v3), hyperthreading, 2.4-GHz maximum frequency, and 132 GB of RAM. While each sub-MS contained identical data, larger datasets \footnote{The original data can be download from \url{https://casa.nrao.edu/Data/EVLA/Pband/P_band_3C129.tgz}, but we extract a 13GB portion of this dataset to work on.} were also tested in our program.
\begin{figure}[h]
 \centering
 \includegraphics[width=\textwidth, angle=0]{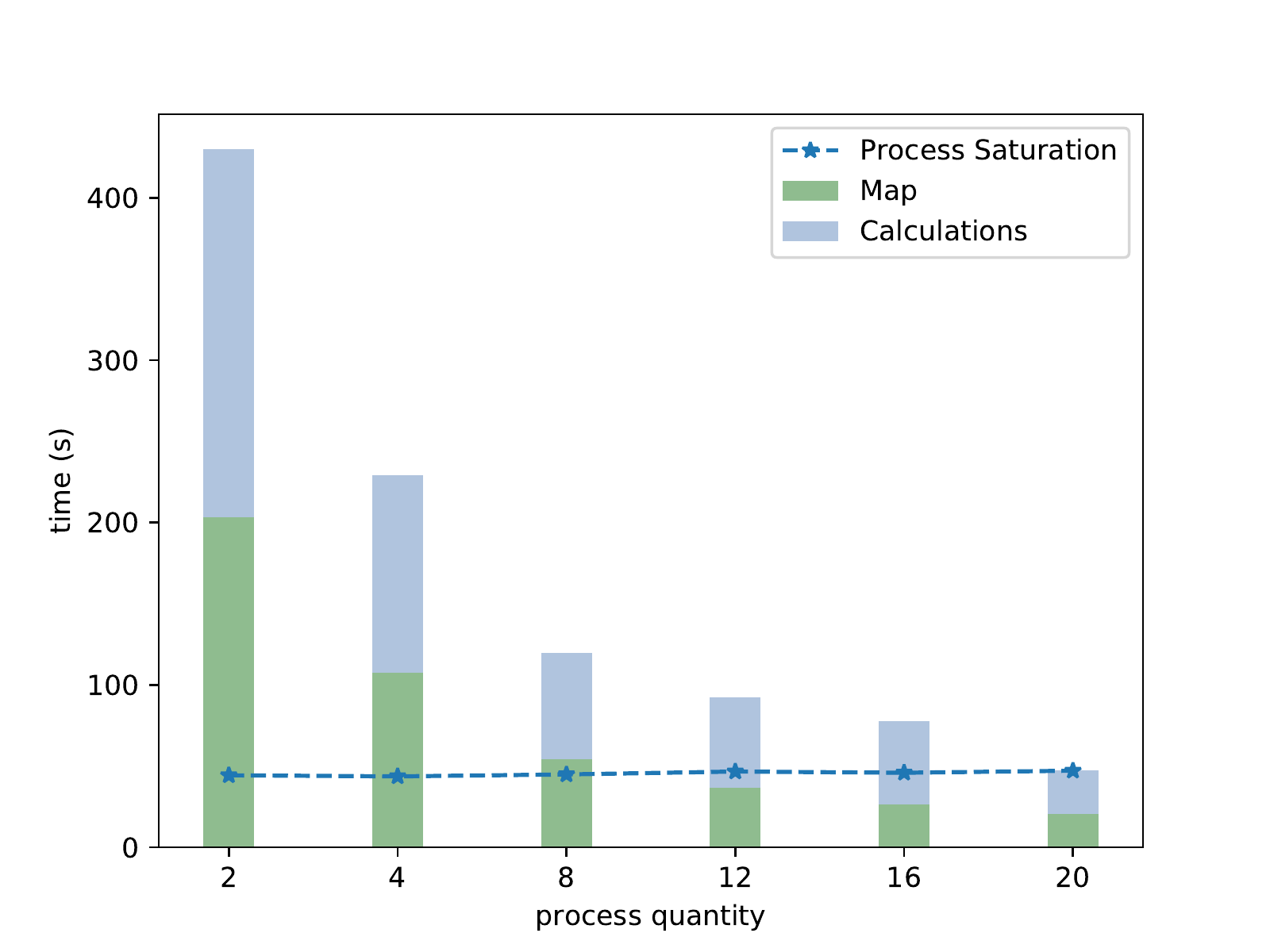}
 \caption{Coarse-grained process execution with identical arguments $t$=2 and $nch$=4 across 20 MSs on 10 nodes (each machine is assigned two dask workers). Process saturation is the number of dask workers, not less than that of MS files, and the execution times indicated by the dashed line do \emph{not} include the reduce phase.}
 \label{ms2020-0112fig4}
\end{figure}

As shown in Figure~\ref{ms2020-0112fig4}, coarse-grained multiprocessing scheduler is able to bypass the GIL issues to stimulate the potential parallelism of multi-cores. Combined with fine-grained \emph{delayed} function, deferred execution can mimic the sequential iteration of the \emph{for-loop} and wrap custom code in sub-operations (see the right in Figure~\ref{fig:ms2020-0112fig2}). Along with the gradual increase in dask workers (or it is equivalent to appending more nodes), the map and calculations execution time decreases significantly, and the minimum value occurs at process saturation. The \textit{Dask} distribution performs well during the loading ($\approx$560MB/s) and averaging phases ($\approx$350MB/s) (marked by the dashed line in Figure~\ref{ms2020-0112fig4}) where time consumption does not increase linearly with the number of the MS. This slow growth is due to the extra coordination required to manage resource competition caused by the large amount of memory accesses and intensive computation. Although distributed environments can increase computing and storage capacity, transferring large amounts of data between processes and separate workstations can introduce more performance penalties. In general, multiprocessing schedulers are an excellent choice when the workflow is relatively linear.

\section{Conclusions}
\label{sect:conclusion}

With the advent of interconnected, fault-tolerant, and resource-sharing clusters, distributed computing is becoming a mainstream method of meeting astronomical computing performance requirements.
This paper presents a fine-grained distributed method via distributed in-memory chunking provided by \textit{Dask} and a labeled data array provided by XArray. Averaging is considered an important approach for reducing the volume of synthetic visibilities. Furthermore, it is advisable to apply maximum possible calibration prior to averaging, as a high CR requires accurate phase calibration. We take averaging as the case and implement a fine-grained distributed averaging over time and channel. The accuracy and efficiency of the proposed method is demonstrated through a series of experiments. Radio interferometric data processing comprises a variety of algorithms that are more sophisticated than averaging, such as calibration, gridding, and imaging, and the parallelization is a consistent theme throughout them. The practicality of fine-grained distributed averaging may also lead to other aspects of interferometric data distributed processing in a future study.

\begin{acknowledgements}
This work is supported by the National Key Research and Development Program of China (2018YFA0404603), the Joint Research Fund in Astronomy (U1831204, U1931141) under cooperative agreement between the National Natural Science Foundation of China (NSFC) and the Chinese Academy of Sciences (CAS), the National Natural Science Foundation of China (No.11903009). Yunnan Key Research and Development Program(2018IA054),Yunnan Applied Basic Research Project (2017FB001, 2018FB103) and Cultivated Program of Kunming University of Science and Technology. The major scientific research project of Guangdong regular institutions of higher learning (2017KZDXM062). This work is also supported by Astronomical Big Data Joint Research Center, co-founded by National Astronomical Observatories, Chinese Academy of Sciences and Alibaba Cloud. The authors wish to thank the reviewers for suggestions that improved the paper.
\end{acknowledgements}

\bibliographystyle{raa}
\bibliography{ms2020-0112}

\begin{thebibliography}{16}
\providecommand\natexlab[1]{#1}
\providecommand\JournalTitle[1]{#1}

\bibitem[Atemkeng {et~al.}(2016)]{atemkeng2016using}
Atemkeng, M., Smirnov, O., Tasse, C., Foster, G., \& Jonas, J. 2016, Monthly
  Notices of the Royal Astronomical Society, 462, 2542

\bibitem[Cornwell {et~al.}(2020)]{RASCIL}
Cornwell, T., Wortmann, P., Nikolic, B., Wang, F., \& Stolyarov, V. 2020, Radio
  Astronomy Simulation, Calibration and Imaging Library,
  \url{https://github.com/SKA-ScienceDataProcessor/rascil}

\bibitem[Cotton(2009)]{cotton2009}
Cotton, W. 2009, Effects of baseline dependent time averaging of UV data

\bibitem[Dewdney {et~al.}(2009)]{dewdney2009square}
Dewdney, P.~E., Hall, P.~J., Schilizzi, R.~T., \& Lazio, T. J.~L. 2009,
  Proceedings of the IEEE, 97, 1482

\bibitem[Hamaker {et~al.}(1996)]{hamaker1996understanding}
Hamaker, J., Bregman, J., \& Sault, R. 1996, Astronomy and Astrophysics
  Supplement Series, 117, 137

\bibitem[Le{\~a}o {et~al.}(2006)]{leao2006circumstellar}
Le{\~a}o, I., De~Laverny, P., M{\'e}karnia, D., De~Medeiros, J., \& Vandame, B.
  2006, Astronomy \& Astrophysics, 455, 187

\bibitem[McMullin {et~al.}(2007)]{mcmullin2007astronomical}
McMullin, J., Waters, B., Schiebel, D., Young, W., \& Golap, K. 2007, RA Shaw,
  F. Hill, \& DJ Bell, 127

\bibitem[Offringa(2016)]{offringa2016compression}
Offringa, A. 2016, Astronomy \& Astrophysics, 595, A99

\bibitem[Perkins {et~al.}(2020)]{daskms}
Perkins, S., Molenaar, G., Smirnov, O., \& Andati, L. 2020, Xarray {D}atasets
  from {C}ASA {T}ables,
  \url{https://dask-ms.readthedocs.io/en/latest/concepts/pyarrays.html}

\bibitem[Raba {et~al.}(2019)]{cngi}
Raba, R., Thomas, A., \& Steeb, J.-W. 2019, CASA {N}ext {G}eneration
  {I}nfrastructure,
  \url{https://cngi-prototype.readthedocs.io/en/0.0.54/visibilities.html}

\bibitem[Rocklin(2015)]{rocklin2015dask}
Rocklin, M. 2015, in Proceedings of the 14th python in science conference No.
  130-136, Citeseer

\bibitem[Smirnov {et~al.}(2012)]{smirnov2012understanding}
Smirnov, O., Frank, B., Theron, I., \& Wood, I.~H. 2012, in 2012 International
  Conference on Electromagnetics in Advanced Applications, IEEE, 586

\bibitem[Smirnov(2011)]{Smirnov2011a}
Smirnov, O.~M. 2011, Astronomy \& Astrophysics, 527, A106

\bibitem[Thompson(1999)]{Thompson1999}
Thompson, A.~R. 1999, in Synthesis Imaging in Radio Astronomy II, Vol. 180, 11

\bibitem[van Diepen(2015)]{van2015casacore}
van Diepen, G. 2015, Astronomy and Computing, 12, 174

\bibitem[Wijnholds {et~al.}(2018)]{wijnholds2018baseline}
Wijnholds, S., Willis, A., \& Salvini, S. 2018, Monthly Notices of the Royal
  Astronomical Society, 476, 2029

\end{thebibliography}

\label{lastpage}

\end{document}